# Experimental demonstration of an ultra-thin three-dimensional thermal cloak

Hongyi Xu[1], Xihang Shi[1], Fei Gao[1], Handong Sun[1,2,*], Baile Zhang[1,2,*]

1. Division of Physics and Applied Physics, School of Physical and Mathematical Sciences, Nanyang Technological University, 21 Nanyang Link, Singapore 637371, Singapore.
2. Centre for Disruptive Photonic Technologies, Nanyang Technological University, 21 Nanyang Link, Singapore 637371, Singapore.

*Electronic mail of corresponding author: hdsun@ntu.edu.sg; blzhang@ntu.edu.sg

## Abstract

We report the first experimental realization of a three-dimensional thermal cloak shielding an air bubble in a bulk metal without disturbing external thermal flux. The cloak is made of a thin layer of homogeneous and isotropic material with specially designed three-dimensional manufacturing. The cloak's thickness is 100 μm while the cloaked air bubble has a diameter of 1 cm, achieving the ratio between dimensions of the cloak and the cloaked object two orders smaller than previous thermal cloaks which were mainly realized in a two-dimensional geometry. This work can find applications in novel thermal devices in the three-dimensional physical space.

An invisibility cloak [1-4] that can hide a three-dimensional (3D) object without disturbing external physical fields must possess a 3D geometry itself. A variety of cloaks working for different physical fields such as electromagnetic fields [5-16], acoustic wave fields [17, 18] and elastic wave fields [19] have been experimentally demonstrated predominantly in a two-dimensional (2D) geometry, where the perceivable visibility in the third dimension was generally ignored. Implementing a 3D cloak is widely acknowledged as being extremely difficult in terms of fabrication and characterization. For example, significant efforts have been made to extend a cloak for electromagnetic waves from 2D to 3D by rotating [10] or extending [11] an original 2D design [20] to fill up the 3D space. However, due to the imperfection of the original 2D design [21], these 3D cloaks are still detectable in principle.

While the majority of cloaking research in the past few years focused on various wave fields, recently diffuse-field cloaks [14-16, 22-24] are attracting more and more attentions. A typical example is the thermal cloak that can hide objects from diffusive heat by guiding thermal energy flux smoothly around a hidden object [25]. Most already reported demonstrations of thermal cloaks were based on transformation thermodynamics [26] (the heat diffusion equation is transformed with a coordinate transformation similar to previous transformation optics) that utilized artificial thermal metamaterials implemented mainly in the 2D geometry [22-24]. A 3D thermal cloak that can hide a 3D object in a thermal environment still remains unrealized. On the other hand, diffusion has distinct properties compared to wave, especially when the diffusion reaches equilibrium and does not change phenomenally. This can be utilized to facilitate the implementation of diffuse-field cloaks. For example, Ref. [15] utilized only commercially available materials to implement a cloak for static magnetic fields without applying coordinate transformation, but adopted a scattering cancellation method similar to previous plasmonic cloaking [2, 9], because at extremely low frequencies the wavelength approaches infinity and local fields behave like diffuse fields reaching the steady state. This ingenious idea apparently will lead to remarkable fundamental breakthrough in cloaking research and significant applications in other relevant research fields that require guiding diffusive field or energy flux.

Here we move one step forward to extend the 2D geometry in Ref. [15] to 3D by transferring the analysis into heat diffusion, and demonstrate the first successful realization of a 3D thermal cloak working in the 3D space. We choose an air bubble as the object to be hidden, since it is well known that stationary air is a poor conductor of heat, and small air gaps can degrade the efficiency of a heat exchange system and cause local over-heating. Moreover, similar to most previous demonstrations of electromagnetic cloaks where a perfect electric conductor (PEC) was shielded to forbid any field penetration, any object put inside the air bubble will also be cloaked from external thermal flux because of the poor thermal conductivity of stationary air. It is worth mentioning that what allows us to successfully fabricate a 3D cloak is a specially designed 3D machining process with that is introduced for the first time in thermal cloaking research, which we will elaborate later.

For clarity purposes, we first briefly introduce the analysis borrowed from Ref. [15] originally for static magnetic field and applied here in heat diffusion. We start with the physical model of the spherical thermal cloak shielding a spherical air bubble in homogenous heat flux in steady state, as illustrated in the inset of Fig. 1. The air bubble, in approximation to a thermal insulator with thermal conductivity $\kappa_1 = 0$, locates at the center with radius $R_1$, surrounded by a single-layer cloak with thermal conductivity $\kappa_2$ and outer radius $R_2$. Homogeneous heat flux diffuses from top to bottom through the background medium with $\kappa_b$. The steady-state thermal diffusion equation in the homogeneous background takes the form as $\nabla^2 T=0$, where $T$ is the temperature distribution in the homogeneous background. Temperatures of the air bubble $T_1$, of the cloak layer $T_2$, and of the background material $T_b$ can be written down as:

$$\begin{cases} T_1 = -\sum_{l=0}^{\infty} A_l r^l P_l(\cos\theta) & (r \leq R_1) \quad (1a) \\ T_2 = -\sum_{l=0}^{\infty} \left[ B_l r^l + C_l r^{-(l+1)} \right] P_l(\cos\theta) & (R_1 < r \leq R_2) \quad (1b) \\ T_b = -\sum_{l=0}^{\infty} \left[ D_l r^l + E_l r^{-(l+1)} \right] P_l(\cos\theta) & (r > R_2) \quad (1c) \end{cases}$$

where $P_l(\cos\theta)$ is the $l$th order Legendre Polynomials, $A_l$ to $D_l$ are constants to be determined, and ($r$, $\varphi$, $\theta$) represent spherical coordinates. With the presence of the thermal cloak, the heat flux is expected to bend around the air bubble and return to its original path after passing the cloak. After solving Eqs. (1) in a similar manner to Ref. [15], we can obtain the thermal conductivity $\kappa_2$ of the cloak as:

$$\kappa_2 = \frac{\kappa_b (2R_2^3 + R_1^3)}{2(R_2^3 - R_1^3)}. \quad (2)$$

To fix the geometry, we plot the dependence of the relative conductivity of the cloak $\kappa_2 / \kappa_b$ on the thickness ratio $R_2/R_1$ in Fig. 1. We assume the background medium is stainless steel with thermal conductivity $\kappa_b = 15$ W/mK. The thermal conductivity of the cloak can be chosen from a variety of materials marked in red circles. Here we choose copper with thermal conductivity $\kappa_2 = 385$ W/mK in both simulation and experiment to achieve a small thickness ratio $R_2/R_1$ at 1.02. Radius $R_1$ and $R_2$ are 0.5cm and 0.51cm, respectively, meaning that a thin layer of copper with thickness of 100 μm is sufficient to cloak an object with diameter of 1 cm.

We perform 3D simulations with the commercial FEM solver COMSOL Multiphysics to verify the effectiveness of the designed thermal cloak in Fig. 2. A heat source with constant temperature $T_1$ at 100 °C contacts the left side of the background medium, and a heat sink with constant temperature $T_2$ at 0 °C contacts the right side of the background medium. The rest boundaries of the simulation domain are defined as insulating boundaries. In Figs. 2a-c, when there is only an idealized air bubble ($\kappa_1 \approx 0$) without the cloak, the distortion of temperature distribution can be observed around the air bubble. In contrast, Figs. 2d-f show that when the cloak is in operation to shield the air bubble, the temperature distortion becomes negligible and all heat flux trajectories are recovered after passing the air bubble. Note that at the beginning of transient state, there is still some moderate distortion around the air bubble (Figs. 2d-e) since the thermal cloak is designed to work ideally in thermal equilibrium. However, in the process of establishing thermal equilibrium, the cloaking function of suppressing temperature distortion is still valid.

The specially designed fabrication process is as follows. Firstly, two blocks of stainless steel ASTM 301 with conductivity of 15 W/mK and dimension of 2×2×1cm were chosen as the background thermal medium where a temperature field can be present. In each block, a hemispherical hole with radius $R_2$= 0.51cm (precision less than 5 µm) was drilled with precise electrical discharge machining (EDM) where a series of rapidly recurring current discharges were applied to precisely erode the hole region in the steel block immersed in a dielectric insulating liquid. The surface roughness of the hemispherical hole was less than 1 µm. Secondly, copper with conductivity 385 W/mK was used to fabricate the single-layer cloak because its excellent ductility is particularly suitable for the fabrication process shown in Fig. 3. As is shown in Fig. 3a, a planar copper disk with thickness of 100 µm and surface area of 1.63cm² (i.e. the surface area of the hemispherical hole in the stainless steel block) was aligned coaxially with the hemispherical hole. A molding rod with a hemispherical tip with radius of 0.5cm was moved exactly above the hole covered by the copper disk. The molding rod then punched the copper disk with punching force around 300 N to press it conformably into the hemispherical hole. Because of the stress between the hemispherical hole in the steel block and the hemispherical tip of the molding rod, a copper shell (Fig. 3b) with homogeneous thickness was formed in the 100μm gap between the hole and the tip. Note that during the process any air defect needs to be avoided to achieve a tight contact between the copper shell and the hole in the steel block. Besides, any misalignment among the hole, the copper disk and the molding rod will also fail the whole manufacturing. Thirdly, two identical half blocks with hemispherical copper shells punched into holes were combined to form a complete 3D cloak, as shown in Fig. 3c. The dimension is 0.5/ 0.51cm for the inner/ outer radius of the copper spherical layer, and 2×2×2cm for the complete stainless steel block.

In the experimental characterization, the top and bottom of the stainless steel block were attached to a hot plate (temperature 100ºC) and an ice tank (temperature 0ºC), respectively (Fig. 3c). The top and bottom of the stainless steel block were coated with Coolermaster® thermal compound to ensure good thermal contact with heat sources. The experiment was conducted in a sealed room to avoid ventilation. Note that the above simulation is under the assumption that stationary air has almost zero thermal conductivity. Practically, stationary air has nonzero thermal conductivity of about 0.02 W/mK, which is close to zero in experiment. Our simulation has verified that this approximation of zero thermal conductivity is valid in experiment. After the system was turned on for about 1 minute, the temperature distribution did not change much. Thereafter the sample was separated into half blocks where the cross-section surface temperature was captured by the infrared (IR) camera FLIR® T640 which was calibrated with an Agema® laboratory black body. Technical challenges in characterization that did not appear in previous experiments of 2D thermal cloaks can be present here. Because the 3D geometry of the spherical cloak actually forms a volume cavity, some resonances of IR light inside the cavity will cause significant artifact for measurement. Moreover, the 3D geometry of the cloak with different depths will cause out-of-focus effect for the IR camera. To correct the error caused by cavity resonances and the out-of-focus effect and exclude the emissivity influence from different materials, a stretched 3M® Scotch Super 33+ Vinyl electrical insulation tape with high emissivity of 0.95 and thickness around 50 µm was attached to the cross-section surface of the steel block whose temperature was to be measured.

Fig. 4 shows the measured temperature distribution and isothermal contours on the cross-section surface. Temperature distortion can be observed around the air bubble without the cloak, as shown in the experimental result in Fig. 4a and the simulated result in Fig. 4b. On the contrary, when the cloak shields the air bubble, the temperature distribution is much less distorted, as shown in the experimental result in Fig. 4c which matches well with the simulation in Fig. 4d. Because the air bubble has nonzero thermal conductivity, its temperature will increase slowly rather than staying constant. This is a common observation in most of previous thermal cloaking experiments up to date. However, the increase of temperature of the air bubble has little influence on the temperature distribution in the external region. Moreover, because of the slow increasing speed of temperature inside the air bubble compared to the surrounding background, this cloak can still provide effective thermal protection for a relatively long time.

In conclusion, we present the first realization of a 3D thermal cloak that can shield a spherical air bubble in a bulk metal from external thermal flux. A special design with concepts borrowed from the previously reported magnetic cloak is adopted without using thermal metamaterials and transformation thermodynamics. Novel 3D fabrication is for the first time introduced to fabricate diffuse-field cloaks. This cloak can find wide cost-effective applications in addressing practical issues of distorted temperature distributions in many relevant industries such as mechanical engine systems and semiconductor devices.

This work was sponsored by Nanyang Technological University under Grants Nos. M4080806 and M4081153, and Singapore Ministry of Education under Grants No. M4011039 and No. MOE2011-T3-1-005.

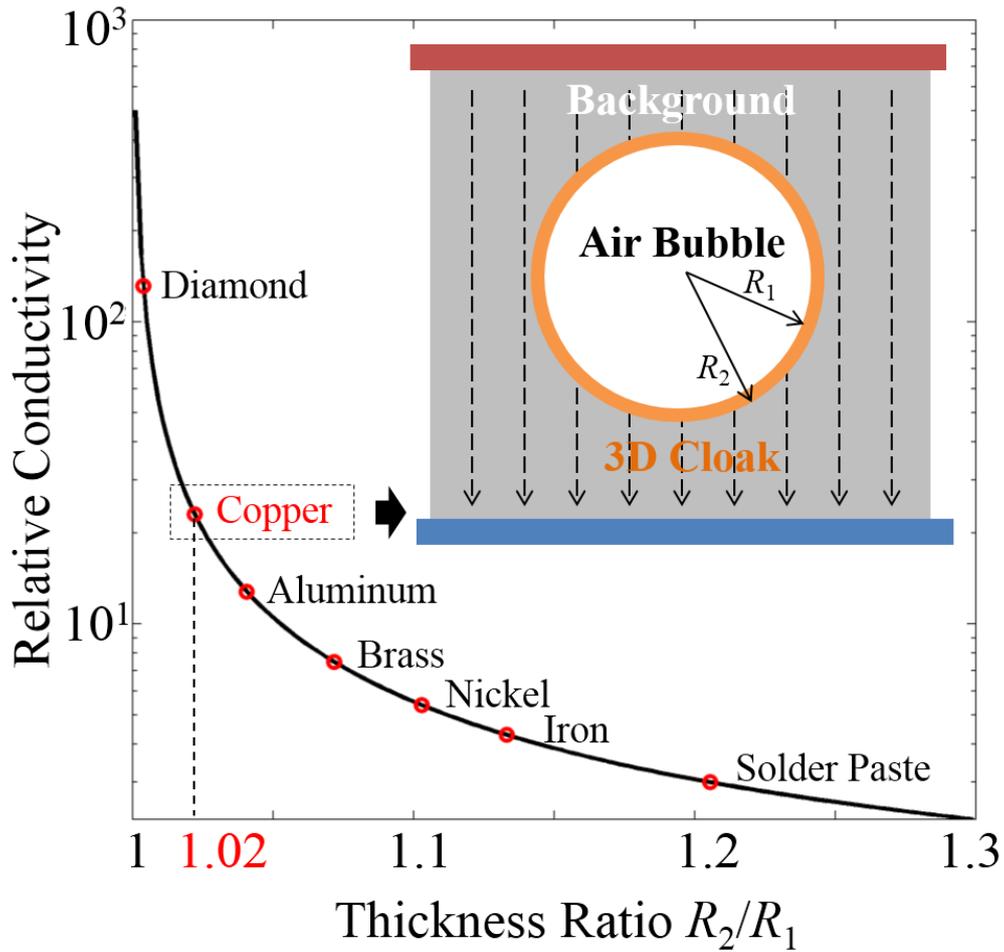

FIG. 1. Material candidates to realize a 3D thermal cloak with the background medium of stainless steel. The black curve shows relative thermal conductivity required to implement a 3D thermal cloak with different thickness ratio of the cloak. The inset figure illustrates the cross section of the cloak (not in scale). Red/blue region denotes high/low temperature, and dashed arrows represent heat flux.

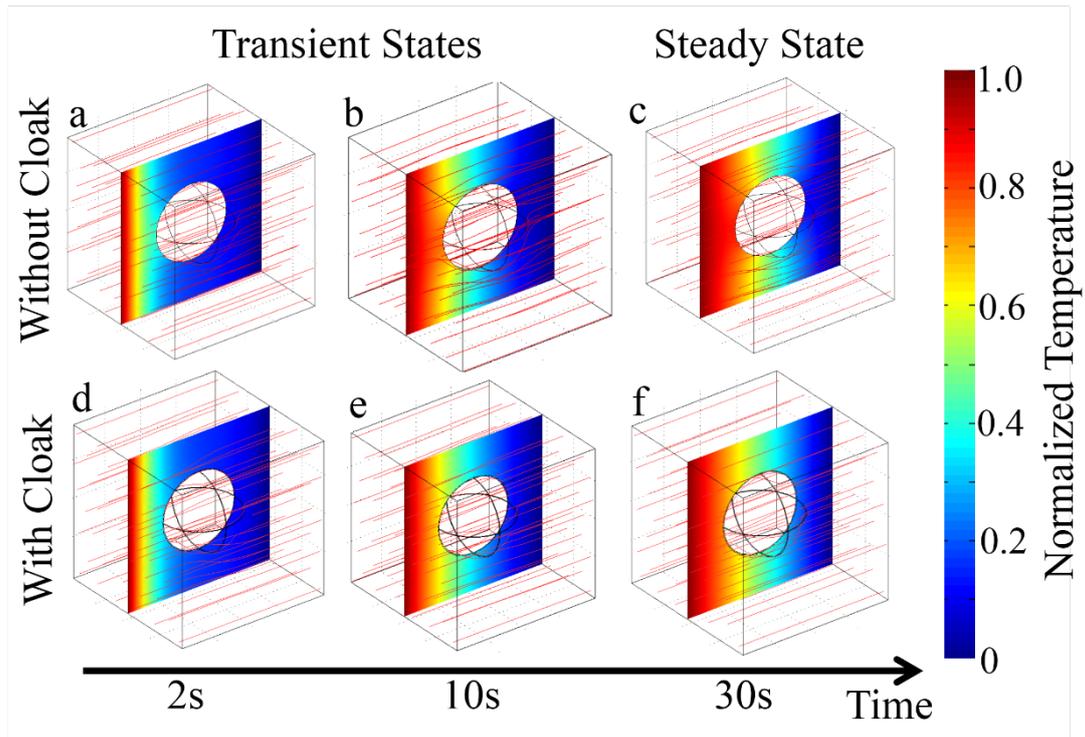

FIG. 2. Simulated temperature distribution and trajectories of heat flux (red curves; intensity of heat flux not represented) for cases without the cloak (a-c) and cases with the cloak (d-f) at different temporal frames 2s, 10s and 30s. Heat flux diffuses from left to right.

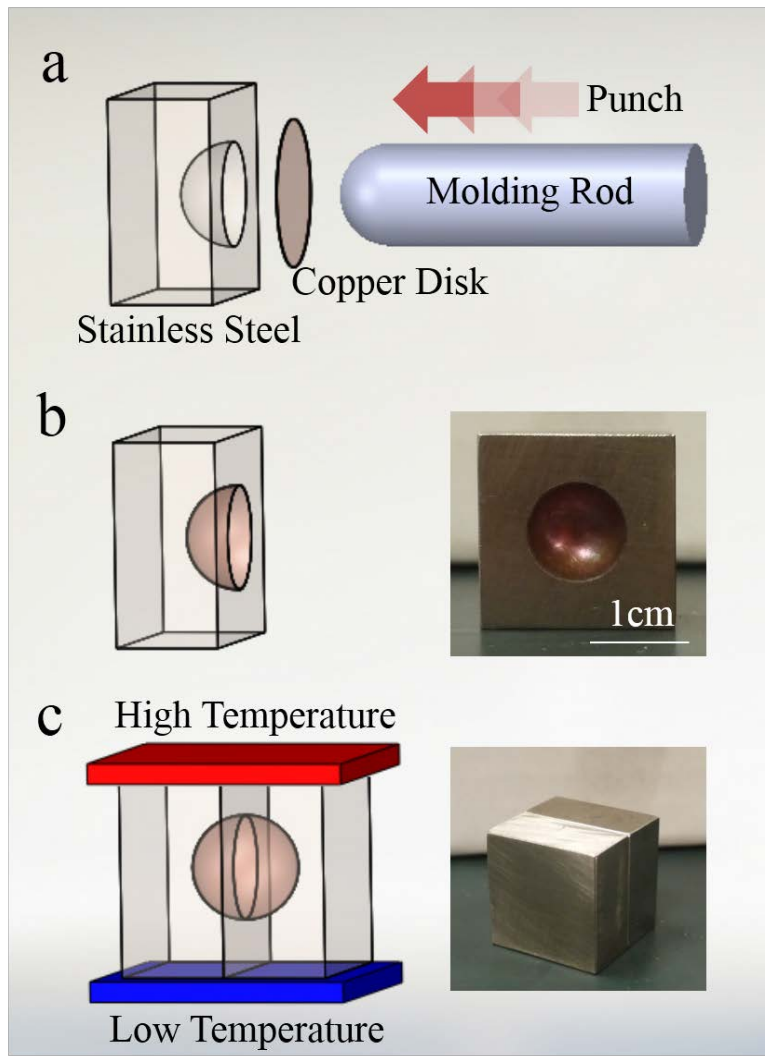

FIG. 3. a. Molding process of half of the 3D thermal cloak: a thin copper disk is punched into the hemispherical hole in the stainless steel block. b. Illustration and snapshot of half of the thermal cloak after molding. c. Illustration and snapshot of the full cloak by combining two half blocks. The red/blue plate represents high/low temperature at the top/bottom surface.

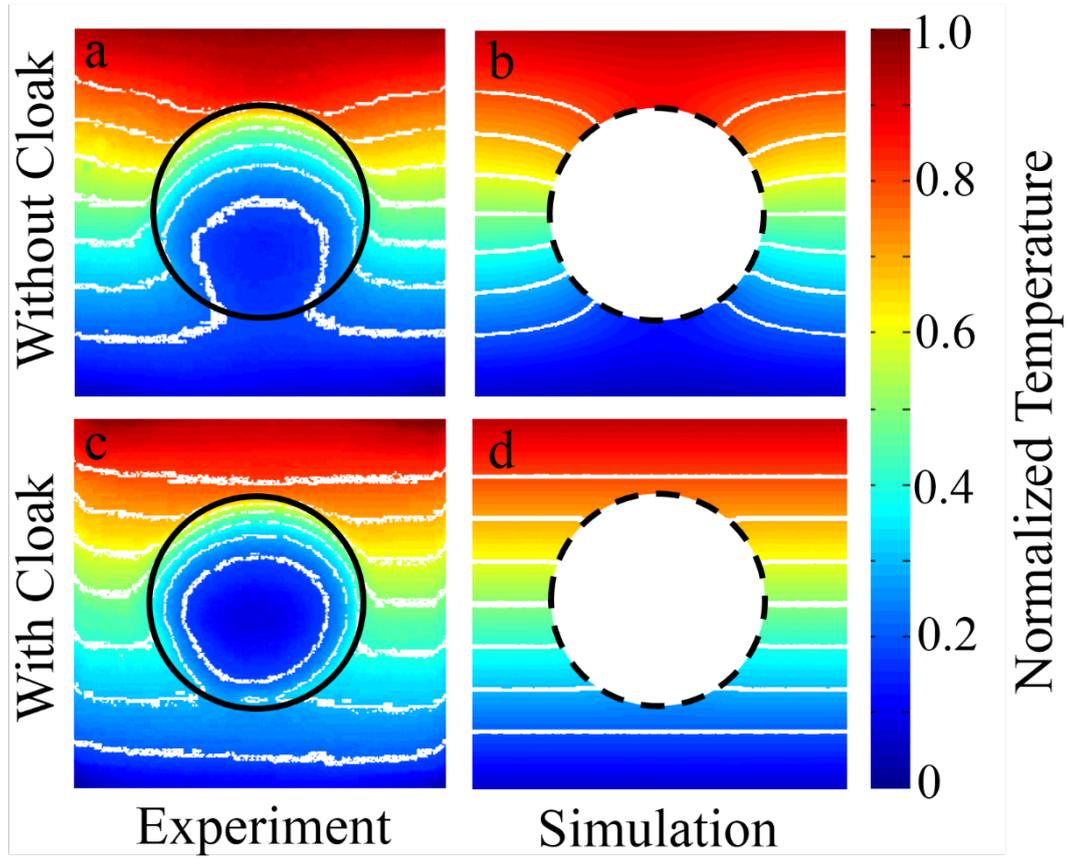

FIG. 4. Distribution and isothermal contours (white curves) of temperature for cases without the 3D thermal cloak (a: experiment and b: simulation) and cases with the 3D thermal cloak (c: experiment and d: simulation).